\documentclass[preprint2]{aastex}
\usepackage{graphicx}
\usepackage{subfigure}

\shorttitle{CLASH: Extending galaxy strong lensing to small physical scales}
\shortauthors{Grillo et al.}

\begin{document}


\title{CLASH: Extending galaxy strong lensing to small physical scales with distant sources highly-magnified by galaxy cluster members$^\star$}


\author{C.~Grillo\altaffilmark{1}, R.~Gobat\altaffilmark{2}, V.~Presotto\altaffilmark{3}, I.~Balestra\altaffilmark{4,5}, A.~Mercurio\altaffilmark{5}, P.~Rosati\altaffilmark{6}, M.~Nonino\altaffilmark{4}, E.~Vanzella\altaffilmark{7}, L.~Christensen\altaffilmark{1}, G.~Graves\altaffilmark{8}, A.~Biviano\altaffilmark{4}, D.~Lemze\altaffilmark{9}, M.~Bartelmann\altaffilmark{10}, N.~Benitez\altaffilmark{11}, R.~Bouwens\altaffilmark{12}, L.~Bradley\altaffilmark{13}, T.~Broadhurst\altaffilmark{14}, D.~Coe\altaffilmark{13}, M.~Donahue\altaffilmark{15}, H.~Ford\altaffilmark{9}, L.~Infante\altaffilmark{16}, S.~Jouvel\altaffilmark{17,18}, D.~Kelson\altaffilmark{19}, A.~Koekemoer\altaffilmark{13}, O.~Lahav\altaffilmark{18}, E.~Medezinski\altaffilmark{9}, P.~Melchior\altaffilmark{20}, M.~Meneghetti\altaffilmark{7}, J.~Merten\altaffilmark{21}, A.~Molino\altaffilmark{11}, A.~Monna\altaffilmark{22}, J.~Moustakas\altaffilmark{23}, L.~A.~Moustakas\altaffilmark{21}, M.~Postman\altaffilmark{13}, S.~Seitz\altaffilmark{22}, K.~Umetsu\altaffilmark{24}, W.~Zheng\altaffilmark{9}, A.~Zitrin\altaffilmark{25,26}}
\email{grillo@dark-cosmology.dk}

\altaffiltext{$\star$}{ This work is based on data collected at NASA HST and at ESO VLT (prog.ID 186.A-0798 and 089.A-0879).}
\altaffiltext{1}{Dark Cosmology Centre, Niels Bohr Institute, University of Copenhagen, Juliane Maries Vej 30, DK-2100 Copenhagen, Denmark}
\altaffiltext{2}{Laboratoire AIM-Paris-Saclay, CEA/DSM-CNRS-Universit\`e Paris Diderot, Irfu/Service d'Astrophysique, CEA Saclay, Orme des Merisiers, F-91191 Gif sur Yvette, France}
\altaffiltext{3}{Dipartimento di Fisica, Universit\`a degli Studi di Trieste, via G. B. Tiepolo 12, I-34143 Trieste, Italy}
\altaffiltext{4}{INAF - Osservatorio Astronomico di Trieste, via G. B. Tiepolo 11, I-34131, Trieste, Italy}
\altaffiltext{5}{INAF - Osservatorio Astronomico di Capodimonte, Via Moiariello 16, I-80131 Napoli, Italy}
\altaffiltext{6}{Dipartimento di Fisica e Scienze della Terra, Universit\`a degli Studi di Ferrara, Via Saragat 1, I-44122 Ferrara, Italy}
\altaffiltext{7}{INAF - Osservatorio Astronomico di Bologna, via Ranzani 1, I-40127 Bologna, Italy}
\altaffiltext{8}{Department of Astrophysical Sciences, Princeton University, Princeton, NJ 08544, USA}
\altaffiltext{9}{Department of Physics \& Astronomy, Johns Hopkins University, 3400 North Charles Street, Baltimore, MD 21218, USA}
\altaffiltext{10}{Institut f\"ur Theoretische Astrophysik, Zentrum f\"ur Astronomie, Universit\"at Heidelberg, Philosophenweg 12, D-69120 Heidelberg, Germany}
\altaffiltext{11}{Instituto de Astrofisica de Andalucia (CSIC), Glorieta de la Astronomia s/n, E-18008 Granada, Spain}
\altaffiltext{12}{Leiden Observatory, Leiden University, P. O. Box 9513, NL-2333 Leiden, The Netherlands}
\altaffiltext{13}{Space Telescope Science Institute, 3700 San Martin Drive, Baltimore, MD 21208, USA}
\altaffiltext{14}{Department of Theoretical Physics, University of the Basque Country, P. O. Box 644, E-48080 Bilbao, Spain}
\altaffiltext{15}{Department of Physics and Astronomy, Michigan State University, East Lansing, MI 48824, USA}
\altaffiltext{16}{Departamento de Astronomia y Astrofisica, Pontificia Universidad Catolica de Chile, V. Mackenna 4860, Santiago 22, Chile}
\altaffiltext{17}{Institut de Cincies de l'Espai (IEE-CSIC), E-08193 Bellaterra (Barcelona), Spain}
\altaffiltext{18}{Department of Physics and Astronomy, University College London, Gower street, London WC1E 6BT, UK}
\altaffiltext{19}{Observatories of the Carnegie Institution of Washington, Pasadena, CA 91101, USA}
\altaffiltext{20}{Center for Cosmology and Astro-Particle Physics, The Ohio State University, 191 W. Woodruff Ave., Columbus, Ohio 43210, USA}
\altaffiltext{21}{Jet Propulsion Laboratory, California Institute of Technology, 4800 Oak Grove Dr, Pasadena, CA 91109, USA}
\altaffiltext{22}{Instituts f\"ur Astronomie und Astrophysik, Universit\"ats-Sternwarte M\"unchen, D-81679 M\"unchen, Germany}
\altaffiltext{23}{Department of Physics and Astronomy, Siena College, 515 Loudon Road, Loudonville, NY 12211, USA}
\altaffiltext{24}{Institute of Astronomy and Astrophysics, Academia Sinica, P.O. Box 23-141, Taipei 10617, Taiwan}
\altaffiltext{25}{Cahill Center for Astronomy and Astrophysics, California Institute of Technology, MS 249-17, Pasadena, CA 91125, USA}
\altaffiltext{26}{Hubble Fellow}


\begin{abstract}

We present a complex strong lensing system in which a double source is imaged five times by two early-type galaxies. We take advantage in this target of the extraordinary multi-band photometric data set obtained as part of the Cluster Lensing And Supernova survey with Hubble (CLASH) program, complemented by the spectroscopic measurements of the VLT/VIMOS and FORS2 follow-up campaign. We use a photometric redshift value of 3.7 for the source and confirm spectroscopically the membership of the two lenses to the galaxy cluster MACS J1206.2$-$0847 at redshift 0.44. We exploit the excellent angular resolution of the HST/ACS images to model the two lenses in terms of singular isothermal sphere profiles and derive robust effective velocity dispersion values of $97 \pm 3$ and $240 \pm 6$ km s$^{-1}$. Interestingly, the total mass distribution of the cluster is also well characterized by using only the local information contained in this lensing system, that is located at a projected distance of more than 300 kpc from the cluster luminosity center. According to our best-fitting lensing and composite stellar population models, the source is magnified by a total factor of 50 and has a luminous mass of approximately $(1.0\pm0.5)\times10^{9}$~$M_{\odot}$ (assuming a Salpeter stellar IMF). By combining the total and luminous mass estimates of the two lenses, we measure luminous over total mass fractions projected within the effective radii of $0.51 \pm 0.21$ and $0.80 \pm 0.32$. Remarkably, with these lenses we can extend the analysis of the mass properties of lens early-type galaxies by factors that are approximately 2 and 3 times smaller than previously done with regard to, respectively, velocity dispersion and luminous mass. The comparison of the total and luminous quantities of our lenses with those of astrophysical objects with different physical scales, like massive early-type galaxies and dwarf spheroidals, reveal the potential of studies of this kind for improving our knowledge about the internal structure of galaxies. These studies, made possible thanks to the CLASH survey, will allow us to go beyond the current limits posed by the available lens samples in the field.

\end{abstract}





\section{Introduction}

Gravitational lensing studies have radically improved our understanding of the internal structure of galaxies and clusters of galaxies (e.g., \citealt{tre10b}; \citealt{bar10}). In particular, the combination of strong lensing with stellar dynamics or stellar population synthesis models has allowed to characterize some properties, previously almost unexplored, of the galaxy dark-matter haloes and sub-haloes. For example, it has become possible to measure the dark over total mass fraction and dark-matter halo density slope in the inner regions of galaxies (e.g., \citealt{gri09,gri10c,gri12}; \citealt{aug09}; \citealt{bar09,bar11}; \citealt{son12}; \citealt{eic12}), and to estimate the mass function of dark satellites (also called substructure) (e.g., \citealt{koc04}; \citealt{veg10,veg12}; \citealt{fad12}; \citealt{xu13}) and the spatial extent (e.g., \citealt{hal07}; \citealt{suy10}; \citealt{ric10}; \citealt{don11}; \citealt{eic13}) of galaxy dark-matter haloes. Moreover, the same combinations of mass diagnostics have enabled to investigate the total mass density profile (e.g., \citealt{rus03,rus05}; \citealt{koo06,koo09}; \citealt{ruf11}; \citealt{son13b}; \citealt{agn13}), the stellar Initial Mass Function (IMF; \citealt{gri08a,gri09}; \citealt{tre10}; \citealt{spi11,spi12}; \citealt{bar13}), and the origin of the tilt of the Fundamental Plane (e.g., \citealt{gri09,gri10b}; \citealt{aug10b}) of massive early-type galaxies and to use single lenses or statistical samples of them to infer cosmologically relevant quantities (e.g., \citealt{gri08b}; \citealt{sch10}; \citealt{suy10b,suy13};  \citealt{fad10}). 

Taking advantage of the excellent data collected by the Lenses Structure and Dynamics (LSD; \citealt{tre04}), Sloan Lens ACS Survey (SLACS; \citealt{bol06}; \citealt{tre06}; \citealt{aug10b}), CFHT Strong Lensing in the Legacy Survey (SL2S; \citealt{mor12}; \citealt{gav12}; \citealt{son13a}), and the BOSS Emission-Line Lens Survey (BELLS; \citealt{bro12}; \citealt{bol12}), the analyses conducted so far have mainly examined the physical properties of isolated, massive early-type galaxies, acting as strong lenses on background sources. Only more recently, thanks also to the Cambridge And Sloan Survey Of Wide ARcs in the skY (CASSOWARY; \citealt{bel09}; \citealt{sta13}), growing interest has been shown in the study of early-type lens galaxies residing in galaxy groups and clusters (e.g., \citealt{gri08c,gri11,gri13}; \citealt{lim09}; \citealt{dea13}). Despite the large amount of results published to date, detailed strong lensing studies in ``small'' lens galaxies are still lacking, mostly because these systems are not observed frequently. One possibility to find more such systems is to look at clusters of galaxies, where, owing to the increase in the strong lensing cross section of a cluster member due to the presence of the extended mass distribution of the cluster, low-mass galaxies are more likely to produce strong lensing features than in less dense environments. Investigations of these objects are particularly useful, as they can provide the necessary piece of information to elucidate what is the amount and distribution of dark matter in astrophysical objects extending from the lowest to the highest ends of the galaxy mass function. The comparison of these observational measurements over a wide range of physical scales with the outcomes of cosmological simulations can give fundamental clues about the precise nature of dark matter and the role played by the interaction of baryons and dark matter during the mass assembly of cosmological structures.

The Cluster Lensing And Supernova survey with Hubble (CLASH; GO 12065, PI Postman) was awarded 524 orbits of Hubble Space Telescope (HST) time to observe 25 massive (virial mass $M_{\mathrm{vir}} \approx 5$-$30 \times 10^{14} M_{\odot}$, X-ray temperature $T_{X} \ge 5$ keV) galaxy clusters in 16 broadband filters, ranging from approximately 2000 to 17000 \AA  $\,$ with the Wide Field Camera 3 (WFC3; \citealt{kim08}) and the Advanced Camera for Surveys (ACS; \citealt{for03}). The sample, spanning a wide redshift range ($z$ = 0.18-0.90), was carefully chosen to be largely free of lensing bias and representative of relaxed clusters, on the basis of their symmetric and smooth X-ray emission profiles (for a thorough overview, see \citealt{pos12}). CLASH has four main scientific goals: 1) measure the cluster total mass profiles over a wide radial range, by means of strong and weak lensing analyses (e.g., \citealt{zit11}; \citealt{coe12}; \citealt{med13}); 2) detect new Type Ia supernovae out to redshift z $\sim$ 2.5, to improve the constraints on the dark energy equation of state (e.g., \citealt{gra13}; \citealt{pat13}); 3) discover and study some of the first galaxies that formed after the Big Bang (z $>$ 7) (e.g., \citealt{zhe12}; \citealt{coe13}; \citealt{bow13}); 4) perform galaxy evolution analyses on cluster members and background galaxies. Ancillary science that can surely be carried out with the superb data set of CLASH is the analysis of several new strong lensing systems on galaxy scale.

A Large Programme (186.A-0798, PI Rosati) of 225 hours with the VIMOS instrument at the Very Large Telescope (VLT) has also been approved to perform a panoramic spectroscopic survey of the 14 CLASH clusters that are visible from ESO-Paranal (Rosati et al. 2014, in preparation). This observational campaign aims at measuring in each cluster the redshifts of 1) approximately 500 cluster members within a radius of more than 3 Mpc; 2) 10-30 lensed multiple images inside the HST field of view, including possible highly-magnified candidates out to z $\approx$ 7 (e.g., \citealt{mon13}; \citealt{bal13}); 3) possible supernova hosts. In one of the CLASH clusters (i.e., MACS J1206.2$-$0847, hereafter MACS 1206), the first spectroscopic redshifts have already been exploited to build robust strong lensing models (\citealt{zit12}; \citealt{ume12}), to obtain an independent total mass estimate from the spatial distribution and kinematics of the cluster members (\citealt{biv13}; \citealt{lem13}), and to confirm a source at $z = 5.703$ (\citealt{bra13}). Strong lensing (with spectroscopically confirmed systems) and cluster dynamics analyses are planned for all 14 southern clusters (e.g., in MACS J0416.1$-$2403, Grillo et al. 2014, in preparation; Balestra et al. 2014, in preparation).

Here, we focus on a rare strong lensing system in which two angularly close early-type galaxies, members of the galaxy cluster MACS 1206 at $z=0.44$, produce in total ten multiple images of a double source located at $z \approx 3.7$. This is the first example of the kind of strong lensing studies that can be conducted on galaxy cluster members, capitalizing on the extraordinary multi-band photometric observations obtained as part of the CLASH program and spectroscopic measurements of the VLT/VIMOS follow-up campaign.

This work is organized as follows. In Sect. 2, we introduce the photometric and spectroscopic observations used for this analysis. In Sect. 3, we present the strong lensing modeling performed to measure principally the total mass values of the lens galaxies and the magnification factors of the multiple images. In Sect. 4, we estimate the luminous mass values of the lens and lensed galaxies by means of stellar population synthesis models. In Sect. 5, we compare some physical quantities, related to the luminous and total masses, of the two lens galaxies with those of lower and higher-mass galaxies. In Sect. 6, we summarize our conclusions. All quoted errors are 68.3\% confidence limits (CL) unless otherwise stated. Throughout this work we assume $H_{0}=70$ km s$^{-1}$ Mpc$^{-1}$, $\Omega_{m}=0.3$, and $\Omega_{\Lambda}=0.7$. In this model, 1\arcsec$\,$ corresponds to a linear size of 5.69 kpc at the cluster redshift of $z=0.44$.

\section{Observations}

\begin{figure}
\centering
\includegraphics[width=0.45\textwidth]{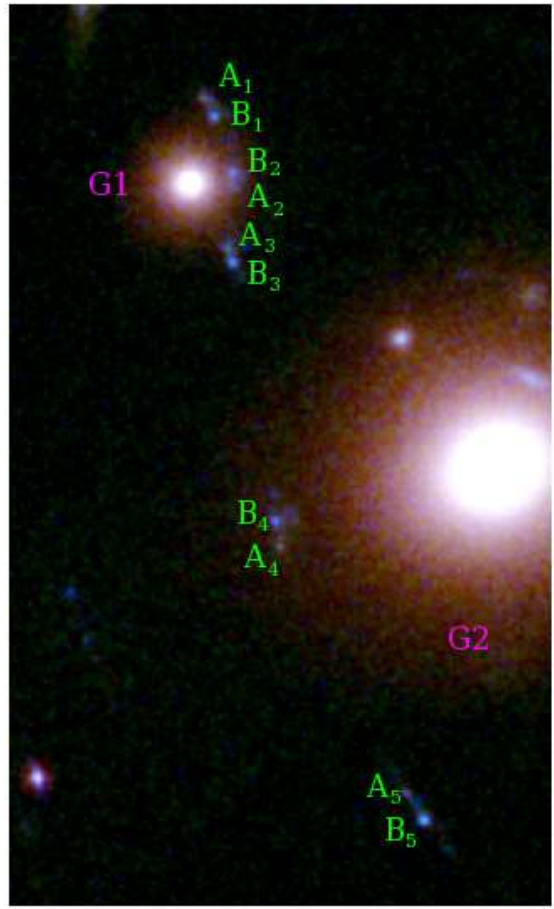}
\caption{Color composite image (9\arcsec$\times$15\arcsec) obtained by combining the F606W+F625W (blue channel), F814W+F850LP (green channel), and F140W+F160W (red channel) filters of HST/ACS and WFC3. Ten multiple images from a double source (at $z \approx 3.7$) are visible around two lens cluster galaxies (at $z=0.44$), G1, near the top, and G2, in the middle. More details on these objects are given in Tables \ref{tab1} and \ref{tab2} and the best-fitting strong lensing model is shown in Figure \ref{fi01}. North is top and East is left.}
\label{fi05}
\end{figure}

\begin{table*}
\centering
\caption{Photometric and spectroscopic properties of the lens galaxies.}
\begin{tabular}{ccccccccccc}
\hline\hline \noalign{\smallskip}
 & R.A. & Decl. & $x$$^{\mathrm{a}}$ & $y$$^{\mathrm{a}}$ & $z_{\mathrm{ph}}$ & $z_{\mathrm{sp}}$ & $q_{L}$ & $\theta_{q_{L}}$$^{\mathrm{b}}$ & $\theta_{e}$ & $n$ \\
 & (J2000) & (J2000) & (\arcsec) & (\arcsec) & & & & (deg) & (\arcsec) & \\
\noalign{\smallskip} \hline \noalign{\smallskip}
G1 & 12:06:16.01 & $-$08:48:17.3 & 0.00 & 0.00 & 0.42 & 0.436 & 0.96 & 113 & 0.43 & 3.7 \\
G2 & 12:06:15.67 & $-$08:48:22.0 & 5.00 & $-$4.68 & 0.50 & 0.439 & 0.84 & 25 & 1.18 & 3.0 \\
\noalign{\smallskip} \hline
\end{tabular}
\begin{list}{}{}
\item[$^{\mathrm{a}}$]With respect to the luminosity center of G1.
\item[$^{\mathrm{b}}$]Angles are measured East of North.
\end{list}
\label{tab1}
\end{table*}

MACS 1206 was observed as part of the CLASH program in HST Cycle 18 between April 3 and July 20 2011 to a total depth of 20 orbits in 16 broadband filters. The images were processed for debias, flats, superflats, and darks using standard techniques, and then co-aligned and combined using drizzle algorithms to a pixel scale of 0.065\arcsec$\,$ (for details, see \citealt{koe07,koe11}). By fitting the full UV to near-IR isophotal aperture magnitudes, photometric redshift estimates of all detected sources were measured through the BPZ (\citealt{ben00,ben04}; \citealt{coe06}) and LePhare (\citealt{arn99}; \citealt{ilb06}) codes.

A color composite image of the lensing system analyzed in this work is shown in Figure \ref{fi05}. The coordinates and photometric redshift measurements, $z_{\mathrm{ph}}$, of the two lenses G1 and G2 and of the multiple images are listed in Tables \ref{tab1} and \ref{tab2}. For the lensed source, we adopt the photometric redshift estimate of 3.7 presented by \citet{zit12} (see system number 7 in the cited paper; for more details on the photometric redshifts measured with the CLASH data, and particularly in MACS 1206, see \citealt{jou13}). This value was determined as an average estimate of the BPZ measurements of the three images (1, 3, and 5; see Figure \ref{fi01}) for which the photometry is less contaminated by the light distribution of the two lens galaxies. In passing, we mention that the BPZ probability distribution functions of the redshift of the three images were all unimodal and that the combined 95\% CL interval of the redshifts extended from 3.4 to 4.2 (this redshift uncertainty is not significant for the lensing analysis performed below; for example, it introduces a percentage error smaller than 1.5\% on the values of the lens effective velocity dispersions plotted in Figure \ref{fi02}). We used the public code \textsc{Galfit}\footnote{http://users.obs.carnegiescience.edu/peng/work/galfit/galfit.html} (\citealt{pen10}) to derive the luminosity structural parameters of the two lenses in the F160W band (see Table \ref{tab1}): the axis ratio, $q_{L}$, the position angle of the major axis, $\theta_{q_{L}}$, the half-light angle, $\theta_{e}$, and the Sersic index, $n$.

Spectroscopic follow-up observations were taken as part of the VLT/VIMOS Large Programme 186.A-0798. Four VIMOS pointings were used, keeping one of the four quadrants fixed to the cluster core in order to allow for long integrations on the strong lensing features. In each pointing, we took 45 to 60 minute exposure times. We used 1\arcsec-wide slits with either the low-resolution LR-Blue grism or the intermediate-resolution MR grism, covering a layout of $20'$-$25'$ across. MACS 1206 is the first of the 14 southern galaxy clusters in the CLASH sample targeted by this spectroscopic program for which the observational campaign was concluded. The measurements have resulted in a total integration time of approximately 11 hours, providing about 600 secure cluster members (see \citealt{biv13}) and 4 confirmed multiple image systems (see \citealt{zit12}). Additional spectroscopic measurements on some cluster members of MACS 1206 were obtained with the VLT/FORS2 in April 17 2012 (Programme ID 089.A-0879, PI Gobat). The observations were taken in good seeing conditions, with the medium resolution grism 600RI and 1\arcsec -wide slits, with a total exposure time of 60 minutes.

\begin{figure*}
\centering
\includegraphics[width=0.84\textwidth]{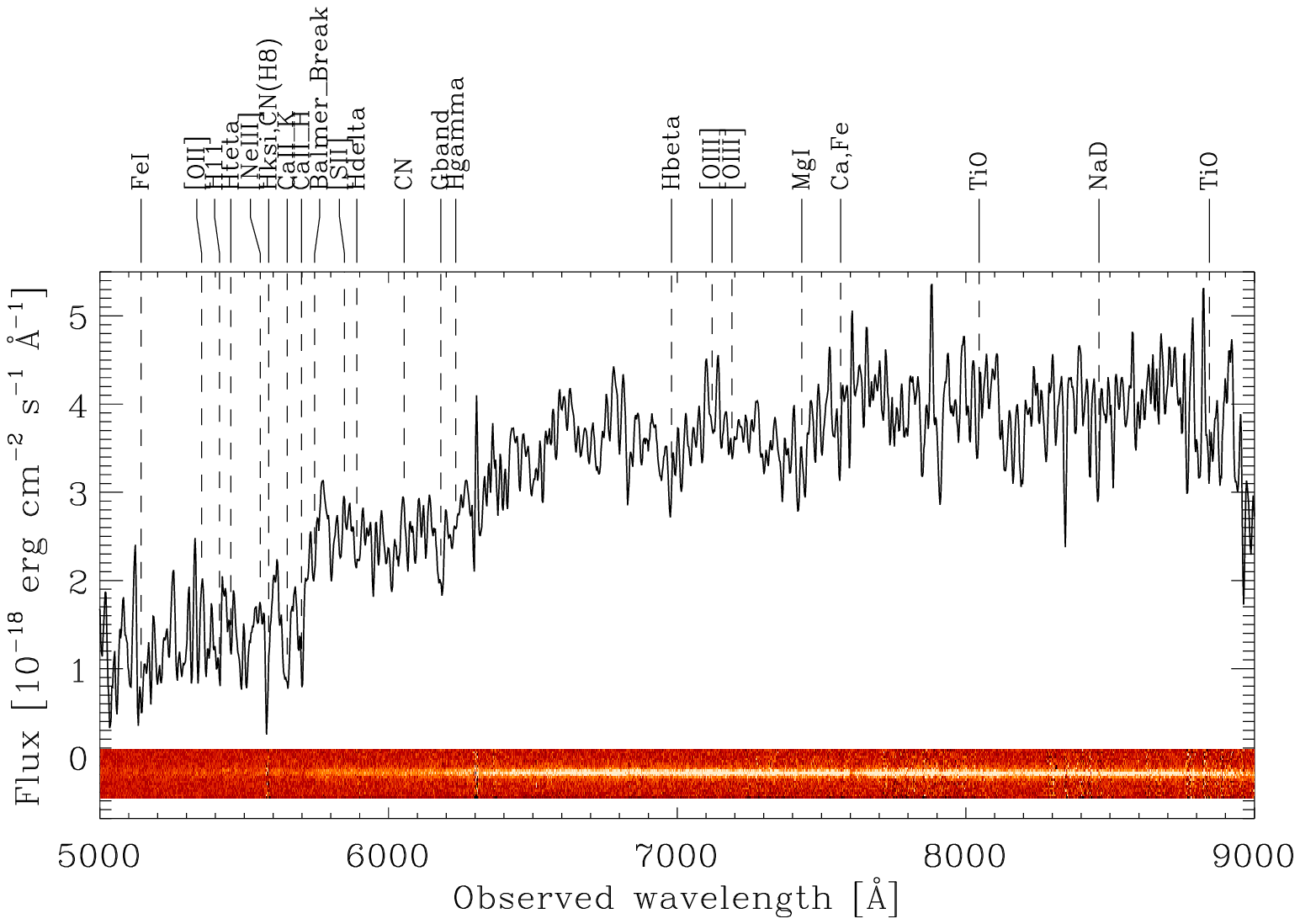}
\includegraphics[width=0.83\textwidth]{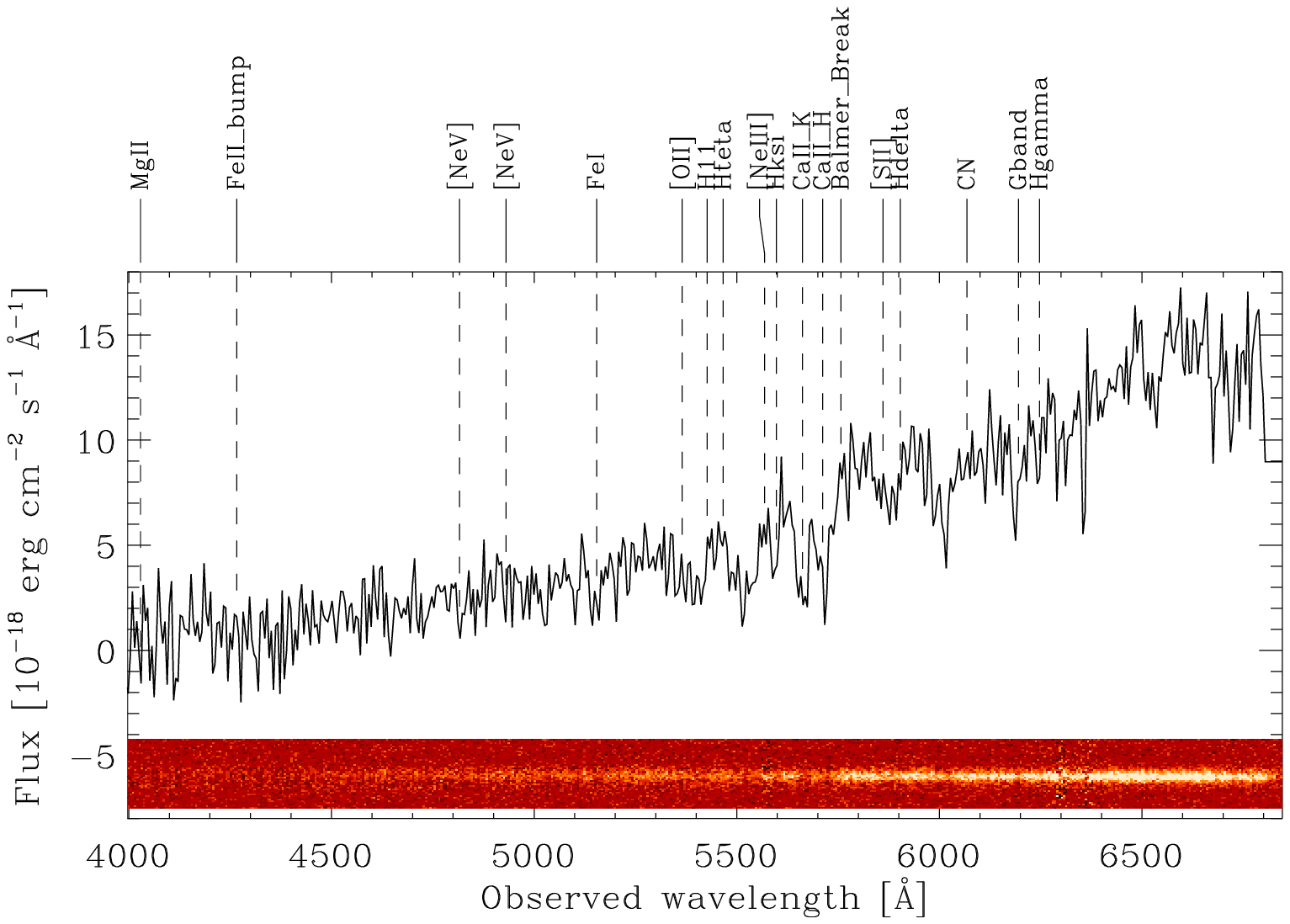}
\caption{Flux-calibrated VLT/VIMOS spectra of G1 and G2 obtained with the MR and LR-Blue grisms, respectively. The exposure times are 45 and 6 minutes. The 1D spectra with several template emission and absorption lines shifted to the best-fitting redshift values and the 2D spectra are shown. \emph{On the top:} Spectra of G1 which provide a redshift value of 0.436. \emph{On the bottom:} Spectra of G2 which provide a redshift value of 0.439.}
\label{fi07}
\end{figure*}

The flux-calibrated VIMOS MR and LR-Blue spectra of G1 and G2, respectively, are shown in Figure \ref{fi07}. The identification of the most prominent absorption lines, like CaII K and H and G-band, have provided spectroscopic redshift estimates, $z_{\mathrm{sp}}$, of 0.436 and 0.439 for these two galaxies. These values are included in the 95\% CL intervals of the estimated photometric redshifts, ranging from 0.36 to 0.46 and from 0.42 to 0.52 for, respectively, G1 and G2. In Figure \ref{fi07}, we also display several template emission and absorption features redshifted to the best-fitting values of the galaxy spectroscopic redshifts. The FORS2 spectrum of G2 is shown in Figure \ref{fi10} and will be discussed in the following section.

\begin{table}[h]
\centering
\caption{Astrometric and photometric measurements for the multiple images.}
\begin{tabular}{ccccc}
\hline\hline \noalign{\smallskip}
 & $x$$^{\mathrm{a}}$ & $y$$^{\mathrm{a}}$ & $z_{\mathrm{ph}}$ & $\delta_{x,y}$ \\
 & (\arcsec) & (\arcsec) & & (\arcsec) \\
\noalign{\smallskip} \hline \noalign{\smallskip}
A$_{1}$ & 0.26 & 1.36 & 3.7 & 0.065 \\
B$_{1}$ & 0.39 & 1.10 & 3.7 & 0.065 \\
B$_{2}$ & 0.72 & 0.20 & 3.7 & 0.065 \\
A$_{2}$ & 0.65 & $-$0.06 & 3.7 & 0.065 \\
A$_{3}$ & 0.58 & $-$1.10 & 3.7 & 0.065 \\
B$_{3}$ & 0.72 & $-$1.30 & 3.7 & 0.065 \\
B$_{4}$ & 1.36 & $-$5.52 & 3.7 & 0.065 \\
A$_{4}$ & 1.50 & $-$5.98 & 3.7 & 0.065 \\
A$_{5}$ & 3.51 & $-$10.01 & 3.7 & 0.065 \\
B$_{5}$ & 3.84 & $-$10.46 & 3.7 & 0.065 \\
\noalign{\smallskip} \hline
\end{tabular}
\begin{list}{}{}
\item[$^{\mathrm{a}}$]With respect to the luminosity center of G1.
\end{list}
\label{tab2}
\end{table}

\section{Strong lensing modeling}

In this section we present two different models of the strong gravitational lensing system. We use the public code \emph{gravlens}\footnote{http://redfive.rutgers.edu/$\sim$keeton/gravlens/} (\citealt{kee01a}) to reconstruct the total mass distribution of the lenses and to estimate the position and magnification factor of the sources. We stress the fact that the analysis presented below has a different perspective compared to the previous strong lensing models of MACS 1206 (\citealt{zit12}; \citealt{eic13}). We concentrate here on radial scales of a few kpc, where the total mass of the galaxies G1 and G2 is the main source of the gravitational potential and the mass of the cluster (extended over a typical radial scale of 100 kpc) is instead approximated and treated as a second-order term.

We keep the model complexity to a minimum and describe the three main lenses, i.e. the two galaxies G1 and G2 and the cluster, in terms of either three singular isothermal spheres (3SISs) or two singular isothermal spheres and a singular isothermal ellipsoid (2SISs+SIE). The mass components are fixed to the luminosity centroids of the galaxies G1 and G2 and of the BCG (the hypothesis of the center of mass of the cluster coinciding with that of the BCG is supported by the results of the studies cited in the previous paragraph) and parametrized by angular scales (labeled as $b_{\mathrm{G1}}$, $b_{\mathrm{G2}}$, and $b_{\mathrm{H}}$), which represent the strength of the lenses and are equal to the values of the Einstein angles in the spherical case. The SIE model requires two additional parameters: the values of the axis ratio, $q$, and of the major axis position angle $\theta_{q}$. The ten multiple images are approximated to point-like objects and associated to two close sources (A and B), each of which is lensed five times. The multiple images are identified with indices running from 1 to 5 (see Table \ref{tab2}). For each image we assume an observational error $\delta_{x,y}$ on the determination of its position of one image pixel (i.e., 0.065\arcsec). Although in each HST filter the luminosity centroids of the multiple images can be measured with positional errors of some fractions of a pixel, we have decided to consider a conservative uncertainty value of one pixel to take into account the point-like approximation, the centroid differences in the individual HST bands, and the contamination from the light of the lens galaxies in estimating the multiple image positions. To quantify the goodness of a model, we use a standard chi-square function, $\chi^{2}$, defined as the sum over all the sources $i$ and their multiple images $j$ of the squared ratios of the differences between the observed ($\bold{x}^{i,j}_{\mathrm{obs}}$) and model-predicted ($\bold{x}^{i,j}_{\mathrm{mod}}$) positions divided by the adopted positional uncertainties ($\sigma_{\bold{x}^{i,j}}$):
\begin{equation}
\chi^{2}(\bold{p}) := \sum_{i,j} \frac{|| \bold{x}^{i,j}_{\mathrm{obs}} - \bold{x}^{i,j}_{\mathrm{mod}} ||^{2}}{\sigma_{\bold{x}^{i,j}}^{2}}\, .
\label{eq:03}
\end{equation}
 We minimize this $\chi^{2}$ estimator by varying the model parameters ($\bold{p}$) and compare the minimum value of the chi-square with the number of degrees of freedom (d.o.f.). These are the number of observables (twenty coordinates of the ten images) minus the number of the parameters of a model (four coordinates for the two sources, three lens angular scales, and, when present, two lens ellipticity parameters). The best-fitting (minimum chi-square) parameters are shown in Table \ref{tab3}. There, we also present the median ($\tilde{\Delta}$) and root mean square ($\Delta_{\mathrm{rms}}$) values of the Euclidean distances ($\Delta$) between the observed and model-predicted angular positions of the multiple images.

\begin{table*}
\centering
\caption{The parameters of the best-fitting models.}
\begin{tabular}{cccccccccc}
\hline\hline \noalign{\smallskip}
Model & $b_{\mathrm{G1}}$$^{a}$  & $b_{\mathrm{G2}}$$^{a}$ & $b_{\mathrm{H}}$ & $q$ & $\theta_{q}$$^{\mathrm{b}}$ & $\chi^2$ & d.o.f. & $\tilde{\Delta}$ & $\Delta_{\mathrm{rms}}$\\
 & (\arcsec) & (\arcsec) & (\arcsec) & & (deg) & & & (\arcsec) & (\arcsec) \\
\noalign{\smallskip} \hline \noalign{\smallskip}
3SISs & 0.21 & 1.25 & 45.3 & & & 19.1 & 13 & 0.081 & 0.090 \\
2SISs+SIE & 0.21 & 1.25 & 37.1$^{c}$ & 0.70 & 14.1 & 13.3 & 11 & 0.059 & 0.075 \\ 
\noalign{\smallskip} \hline
\end{tabular}
\begin{list}{}{}
\item[$^{\mathrm{a}}$]See Eq. (\ref{eq:02}) for the interpretation of the lens strength $b$ in terms of effective stellar velocity dispersion $\sigma$.
\item[$^{\mathrm{b}}$]Angles are measured East of North.
\item[$^{\mathrm{c}}$]Note that Gravlens provides the value of the lens strength $b$ multiplied by a function $f(\cdot)$ of the minor-to-major axis ratio, $q$ (see \citealt{kee01a,kee01b}).
\end{list}
\label{tab3}
\end{table*}

The values of the best-fitting chi-square are very close to the number of degrees of freedom. This fact and the small values of $\tilde{\Delta}$ and $\Delta_{\mathrm{rms}}$ confirm that our relatively simple mass modeling choices are adequate to describe the lenses. More quantitatively, the probability that the value of a random variable extracted from the chi-square distribution with 11(13) degrees of freedom is greater than 13.3(19.1) is 0.27(0.12). It is not surprising that the spherical approximation used for describing the total mass distributions of the galaxies G1 and G2 is proved suitable, given the large values of the minor-to-major axis ratio $q_{L}$ of their luminous components (see Table \ref{tab1}). Looking at Table \ref{tab3}, we conclude that the values of $b_{\mathrm{G1}}$ and $b_{\mathrm{G2}}$, which characterize the total mass distribution of the two main lenses, are robust and not sensitive to the details of the cluster total mass modeling.

\begin{figure*}
\centering
\includegraphics[width=0.45\textwidth]{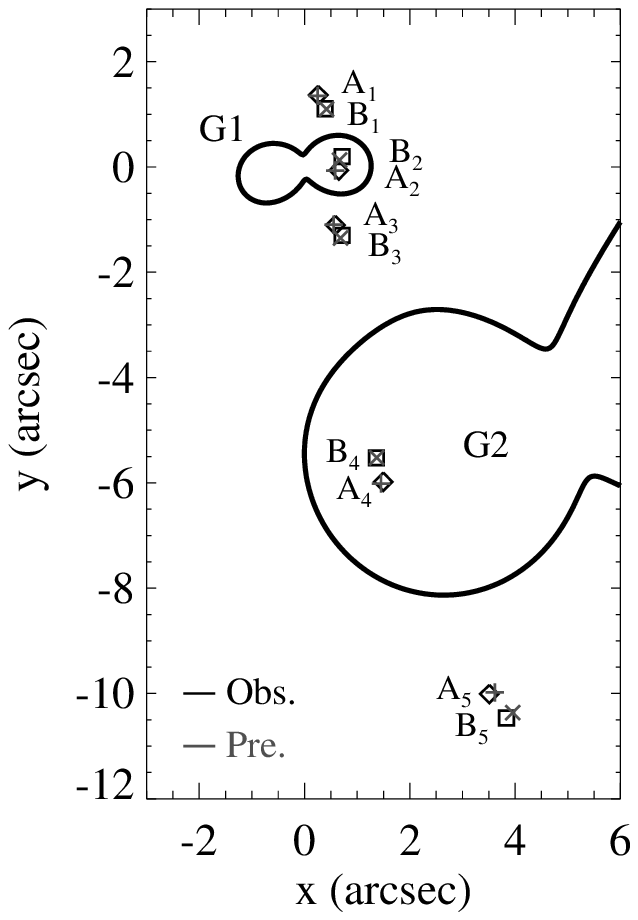}
\includegraphics[width=0.45\textwidth]{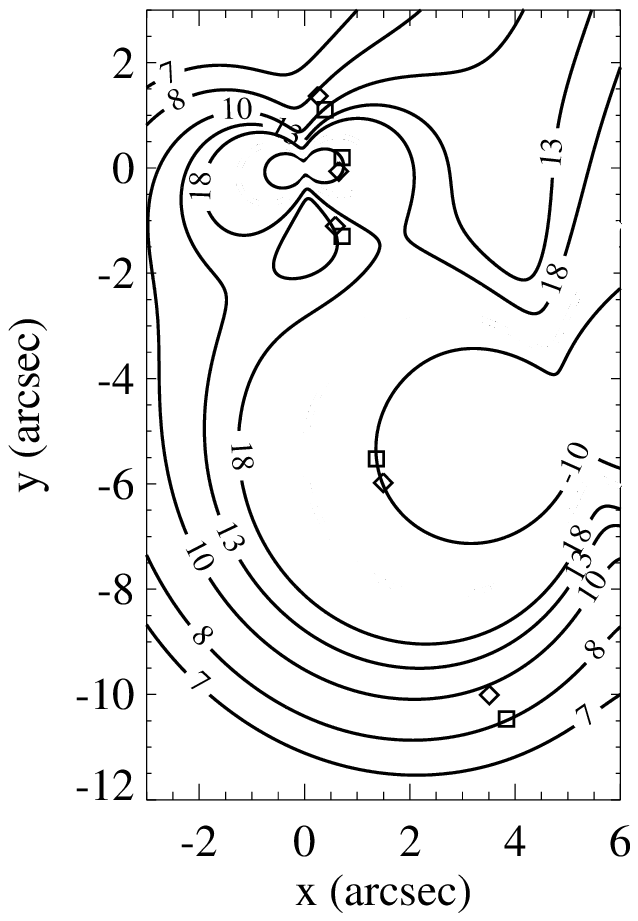}
\caption{The best-fitting 2SISs+SIE strong lensing model. \emph{On the left:} The model-predicted critical curves and multiple image positions (diamond and square symbols) of the two sources, $A$ and $B$, around the two main lenses, G1 and G2. The observed positions of the multiple images (cross and plus symbols) are shown for comparison. \emph{On the right:} Reconstructed values of the magnification factor at the positions of the model-predicted multiple images. Positive and negative values on the contour levels indicate, respectively, if the images have conserved or inverted parity with respect to the source. North is top and East is left.}
\label{fi01}
\end{figure*}

In Figure \ref{fi01}, we plot the best-fitting 2SISs+SIE model with the observed and model-predicted multiple images and the critical curves. We show also the values of the magnification factors in proximity to the reconstructed positions of each multiple image.

Interestingly, starting from the only lensing system analyzed in this work we measure that for the cluster component, when modeled as an SIE, the values of axis ratio and position angle are aligned with the prominent intracluster light (for more details on the properties of the intracluster light in MACS 1206, we refer to Presotto et al. 2014, in preparation). The values of these parameters are consistent with those obtained from the thorough lensing analyses performed on the cluster scale by \citet{zit12}, \citet{eic13}, and \citet{ume12}. The first two and the last studies exploit, respectively, the full strong lensing and strong plus weak lensing information in MACS 1206. Moreover, the best-fitting parameters for the SIE model associated to the cluster component provide a total mass estimate of about $4\times10^{14} M_{\odot}$ projected within a cylinder of radius equal to 320 kpc (the approximate average distance of the multiple images from the BCG luminosity center). Given our simplified assumptions on the cluster total mass distribution, it is remarkable that this estimate is only approximately $10\%$ higher than those obtained in the previously cited lensing works and in the cluster dynamical analysis by \citet{biv13}. If one used otherwise the very crude approximation of the Einstein radius of the cluster (for a source at redshift 3.7) given by the average projected distance between the strong lensing system and the BCG luminosity centre, this would translate into a total projected mass that is more than 1.5 times larger than what obtained from the other cluster total mass diagnostics. Several previous studies of strong lensing systems around galaxy cluster members (e.g., \citealt{gri08c}; \citealt{lim09}) have demonstrated that although these systems contain enough information to characterize reasonably well the cluster mass distribution, this last term is not the main focus of such studies and it can be safely modeled with an approximated convergence plus shear contribution.

In Table \ref{tab5}, we consider the two best-fitting models and list the values of the magnification factor for images A$_{5}$ and B$_{5}$, $\mu(\mathrm{A}_{5})$ and $\mu(\mathrm{B}_{5})$, and of the total magnification factor for the sources A and B, i.e., the sum of the magnification values over all the multiple images of each source:
\begin{eqnarray}
\mu_{\mathrm{tot}}(\mathrm{A}) &:=& \sum_{i=1}^{5} \mu(\mathrm{A}_{i}) \, ; \nonumber \\
\mu_{\mathrm{tot}}(\mathrm{B}) &:=& \sum_{i=1}^{5} \mu(\mathrm{B}_{i}) \, .
\label{eq:01}
\end{eqnarray}
We decide to concentrate on images A$_{5}$ and B$_{5}$ because they are the most distant objects from the luminosity (and mass) centers of the two lenses G1 and G2. For this reason, their photometry is less contaminated by the light distribution of the much brighter lens galaxies. In addition, their relatively large distance from the critical curves (see Figure \ref{fi01}) makes the measurements of their magnification factors less dependent on the modeling details. We conclude that A$_{5}$ and B$_{5}$ are magnified by a factor of approximately 9 and that each of the two sources is magnified in total by a factor of approximately 50.

\begin{table}
\centering
\caption{Values of the magnification factor for the best-fitting models.}
\begin{tabular}{ccccc}
\hline\hline \noalign{\smallskip}
Model & $\mu(\mathrm{A}_{5})$ & $\mu(\mathrm{B}_{5})$ & $\mu_{\mathrm{tot}}(\mathrm{A})$ & $\mu_{\mathrm{tot}}(\mathrm{B})$ \\
\noalign{\smallskip} \hline \noalign{\smallskip}
3SISs & 9.8 (0.4) & 8.5 (0.3) & 48 (1.2) & 50 (1.4) \\
2SISs+SIE & 9.3 (0.4) & 8.1 (0.3) & 47 (1.3) & 50 (1.5) \\ 
\noalign{\smallskip} \hline
\end{tabular}
\begin{list}{}{}
\item[Notes --]In parentheses, we show the 1$\sigma$ statistical uncertainties derived from the bootstrapping analysis.
\end{list}
\label{tab5}
\end{table}

\begin{figure}
\centering
\includegraphics[width=0.44\textwidth]{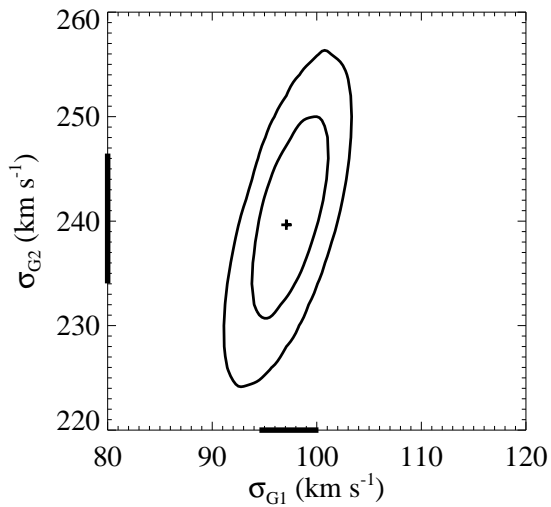}
\caption{Values of the effective velocity dispersion of the two main lenses, G1 and G2, for the 2SISs+SIE model. The best-fitting values are represented by a cross. The 68\% and 95\% confidence regions (contour levels) and the 68\% confidence intervals (thick lines on the axes) are obtained from the $\chi^{2}$ minimization of $10^{4}$ resampled multiple image positions.}
\label{fi02}
\end{figure}

To estimate the statistical uncertainties on the model parameters $b_{\mathrm{G1}}$ and $b_{\mathrm{G2}}$, we perform a bootstrapping analysis in the 2SISs+SIE case. We resample the position of the 10 multiple images by extracting random values from Gaussian distributions with average and standard deviation values equal to, respectively, the positions and positional uncertainties listed in Table \ref{tab2}. We simulate in this way $10^{4}$ data samples, minimize the positional $\chi^{2}$ shown in Eq. (\ref{eq:03}), and consider the best-fitting values of the lens strength. We recall that the value of the effective velocity dispersion $\sigma$ of an SIS model is related to that of the lens strength $b$ in the following way:
\begin{equation}
b = 4 \pi \bigg( \frac{\sigma}{c} \bigg)^2 \frac{D_{ls}}{D_{os}} \, ,
\label{eq:02}
\end{equation}
where $c$ is the speed of light and $D_{ls}$ and $D_{os}$ are, respectively, the angular diameter distances between the lens and the source and the observer and the source. We show the results of this analysis in Table \ref{tab4} and Figure \ref{fi02}. 

\begin{table}
\centering
\caption{Median and statistical 1$\sigma$ error values of the SIS effective velocity dispersions of the two main lenses.}
\begin{tabular}{ccc}
\hline\hline \noalign{\smallskip}
Model & $\sigma_{\mathrm{G1}}$ & $\sigma_{\mathrm{G2}}$ \\
 & (km s$^{-1}$) & (km s$^{-1}$) \\
\noalign{\smallskip} \hline \noalign{\smallskip}
3SISs & $100 \pm 3$ & $241 \pm 6$ \\
2SISs+SIE & $97 \pm 3$ & $240 \pm 6$ \\ 
\noalign{\smallskip} \hline
\end{tabular}
\label{tab4}
\end{table}

\begin{figure}[h]
\centering
\includegraphics[width=0.47\textwidth]{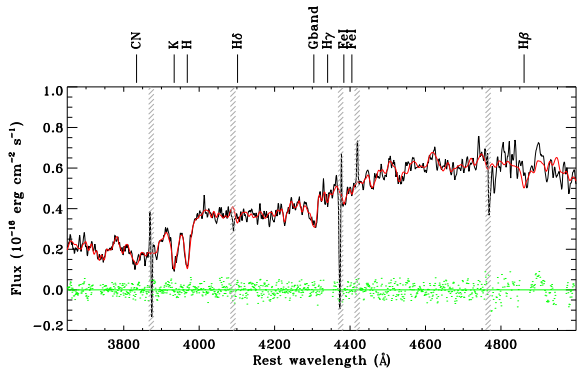}
\caption{VLT/FORS2 rest-frame spectrum of G2 from an aperture of 6 pixels (i.e., 1.5 arcsec) around the peak of emission. The total exposure time is 60 minutes. The data have been slightly smoothed with a box of 3 pixels for display purpose. Black and red solid lines show the data and the best-fitting model, respectively, and green dots display the difference of these last two spectra. The grey shaded areas correspond to the sky line regions which have been masked while performing the fit due to their high residuals. The main absorption lines and possible emission lines are labelled.}
\label{fi10}
\end{figure}

The values of $\sigma_{\mathrm{G1}}$ and $\sigma_{\mathrm{G2}}$ are positively correlated and measured with a small statistical uncertainty. The median values with 1$\sigma$ errors of $\sigma_{\mathrm{G1}}$ and $\sigma_{\mathrm{G2}}$ are $97 \pm 3$ and $240 \pm 6$ km s$^{-1}$, respectively. As already found in several strong lensing studies (e.g., \citealt{gri08c,gri11}), the total mass projected within the average distance of the multiple images from the lens center can be measured precisely. In this specific system, we confirm that an increase of the total mass component associated to the two lens galaxies (i.e., larger values of $\sigma_{\mathrm{G1}}$ and $\sigma_{\mathrm{G2}}$) is correlated to a decrease of the total mass contribution related to the cluster (i.e., smaller values of $\sigma_{\mathrm{H}}$), and viceversa, in order to keep their sum approximately constant. 

Then, we repeat two more times the bootstrapping analysis in the 2SISs+SIE case, first allowing the center of the SIE component (i.e., the cluster dark-matter halo) to vary and then including the two candidate (based on the photometric redshift values) cluster members nearest in projection to G1 and G2. In the first case, we obtain $94 \pm 3$ and $227 \pm 8$ km s$^{-1}$ for the median and 1$\sigma$ error values of the SIS effective velocity dispersions of the two main lenses, $\sigma_{\mathrm{G1}}$ and $\sigma_{\mathrm{G2}}$, respectively. Compared to the previous estimates (see the last row in Table \ref{tab4}), we remark that our assumption on the center of mass of the cluster dark-matter halo does not affect significantly the measurements of the most relevant quantities of our lensing model, i.e. $\sigma_{\mathrm{G1}}$ and $\sigma_{\mathrm{G2}}$. Furthermore, the estimates of the lens strength of the halo, $b_{\mathrm{H}}$, change from $37 \pm 1$\arcsec$\,$ (in the previous analysis with the total mass center fixed) to $38 \pm 2$\arcsec$\,$ (in this new analysis with the total mass center free). We remind that the squared value of the strength of a lens is proportional to the mass of that lens projected within its Einstein radius. From this consideration and from the cited estimates of $b_{\mathrm{H}}$, we notice that the mass measurements of the cluster dark-matter halo with and without its mass center fixed are consistent, given their uncertainties. In the second case, we add two galaxies at a projected distance from G1 of approximately 6\arcsec$\,$ (to the North), with luminosity and size values not larger than those of G1. We include these two lenses in the model, fixing their total mass center and strength values (the former to the galaxy luminosity centroids and the latter to the upper limit given by the strength value of G1), and find $97 \pm 3$ and $249 \pm 6$ km s$^{-1}$, respectively, for $\sigma_{\mathrm{G1}}$ and $\sigma_{\mathrm{G2}}$. As from the previous test, comparing these new values with those of Table \ref{tab4}, we can exclude a significant effect of the two nearest candidate cluster members on our estimates of the total mass distributions of G1 and G2. Given their larger projected distances, the possible influence of other cluster members is expected to be even smaller that that of the two neighboring galaxies considered above. From these tests, we can confirm that our measurements of $\sigma_{\mathrm{G1}}$ and $\sigma_{\mathrm{G2}}$ are robust and that the possible systematic uncertainties, due to our specific modeling assumptions, are approximately on the same order of the statistical uncertainties. This means that, even considering both statistical and systematic errors, the values of $\sigma_{\mathrm{G1}}$ and $\sigma_{\mathrm{G2}}$ can be measured with relative errors of less than 10\% and, therefore, that the errors on these quantities are not dominating the error budget of the galaxy luminous over total mass fractions presented in Sect. 5. The errors on these last quantities are in fact mainly driven by the errors on the luminous mass values, estimated in the next section from the galaxy spectral energy distribution fitting.

We remark that our estimates of $\sigma_{\mathrm{G1}}$ and $\sigma_{\mathrm{G2}}$ are consistent, given the errors, with the values of 101 and 236 km s$^{-1}$, respectively, obtained by \citet{eic13}. They performed a strong lensing study of this cluster using the multiple images of 13 background sources and modeling the cluster total mass distribution with a combination of an extended NFW profile and several, smaller, truncated isothermal profiles (representing the candidate cluster member mass contribution), scaled according to the Faber-Jackson relation (\citealt{fab76}). We emphasize that the adoption of scaling relations to model the total mass distributions of candidate cluster members, necessary in order to reduce the number of parameters of a cluster strong lensing model, provides interesting results on the statistical ensemble of galaxies, but these results should be interpreted very carefully if the main focus is on the study of the mass properties of individual cluster members. We caution that the choice of particular scaling relations can drive the results on possible variations in the amount of dark matter present in the inner regions of different cluster members. For this reason, tailored strong lensing models, like the one presented above, are needed.

Furthermore, we measure the stellar velocity dispersion of G2, $\sigma_{*,\mathrm{G2}}$, within an aperture with diameter equal to 1.5\arcsec$\,$, from a fit of the VLT/FORS2 spectrum shown in Figure \ref{fi10}. We use the pixel-fitting method of \citet{cap04} and adopt the template stellar spectra of the MILES library (\citealt{san06,fal11}). We estimate a redshift of 0.4402, consistent with the VIMOS measurement, and a value of $\sigma_{*,\mathrm{G2}}$ of $(250 \pm 30)$ km s$^{-1}$, also consistent, given the uncertainties, with our strong lensing estimate. We notice that a good agreement between the values of the effective velocity dispersion of an isothermal model and of the central stellar velocity dispersion is common in galaxy-scale strong lensing systems where the multiple image geometry can be reconstructed well (e.g., \citealt{tre06,gri08b}).

By exploiting the optimized lens mass models obtained from our bootstrapping analysis, we can also estimate the statistical uncertainties on the values of the different magnification factors. We show the 1$\sigma$ errors (in parentheses) in Table \ref{tab5}. The values of only a few per cent for the magnification relative errors are not surprising, because of the large number of multiple images that provide detailed information about the lens total mass distributions. We remark that the small errors on the total magnification factors are a minor source of uncertainty on the unlensed luminous mass of the source presented in the next section.

Finally, assuming that an isothermal profile is a good description of the total mass distribution of the two lenses out to their effective radius (for these values in angular units, see $\theta_{e}$ in Table \ref{tab1}) and using the results of the bootstrapping analysis (see Table \ref{tab4}), we measure total mass values $M_{T}$ projected within $R_{e}$,
\begin{equation}
M_{T}(<R_{e}) = \frac{\pi \sigma^{2} R_{e}}{G} \, ,
\label{eq:04}
\end{equation}
(being $G$ the value of the gravitational constant) of $1.7_{-0.1}^{+0.1}\times10^{10}$ and $2.8_{-0.1}^{+0.2}\times10^{11}$ $M_{\odot}$ for $M_{T,\mathrm{G1}}(<2.4\, \mathrm{kpc})$ and $M_{T,\mathrm{G2}}(<6.7\, \mathrm{kpc})$, respectively.  

\begin{table*}
\centering
\caption{Details of the composite stellar population models adopted to measure the luminous mass values of G1, G2, and $\mathrm{A}_{5}+\mathrm{B}_{5}$ (after correcting for the lensing magnification effect).}
\begin{tabular}{cccccc}
\hline\hline \noalign{\smallskip}
 & IMF & SFH & $Z$ & dust & $M_{L}$ \\
\noalign{\smallskip} \hline \noalign{\smallskip}
G1 & Salpeter & delayed exponential & $Z_{\odot}$ & yes & $(1.7 \pm 0.7)\times10^{10}$ $M_{\odot}$ \\
G2 & Salpeter & delayed exponential & $Z_{\odot}$ & yes & $(4.5 \pm 1.8)\times10^{11}$ $M_{\odot}$ \\
$\mathrm{A}_{5}+\mathrm{B}_{5}$ & Salpeter & constant & $Z_{\odot}$ & yes & $(1.0 \pm 0.5)\times10^{9}$ $M_{\odot}$ \\
\noalign{\smallskip} \hline
\end{tabular}
\label{tab7}
\end{table*}

\section{Luminous mass estimates}

\begin{figure*}
\centering
\includegraphics[width=0.49\textwidth]{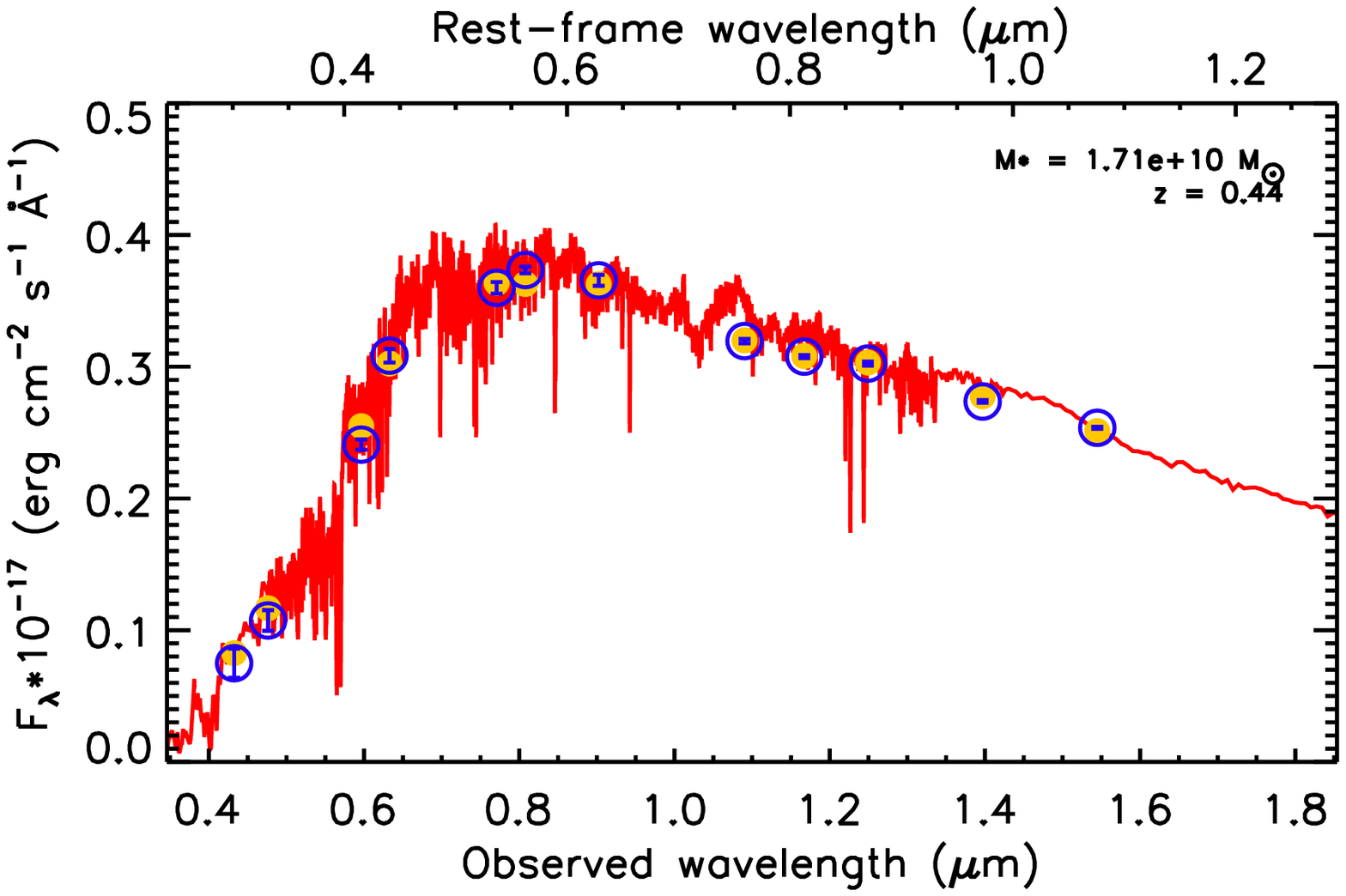}
\includegraphics[width=0.49\textwidth]{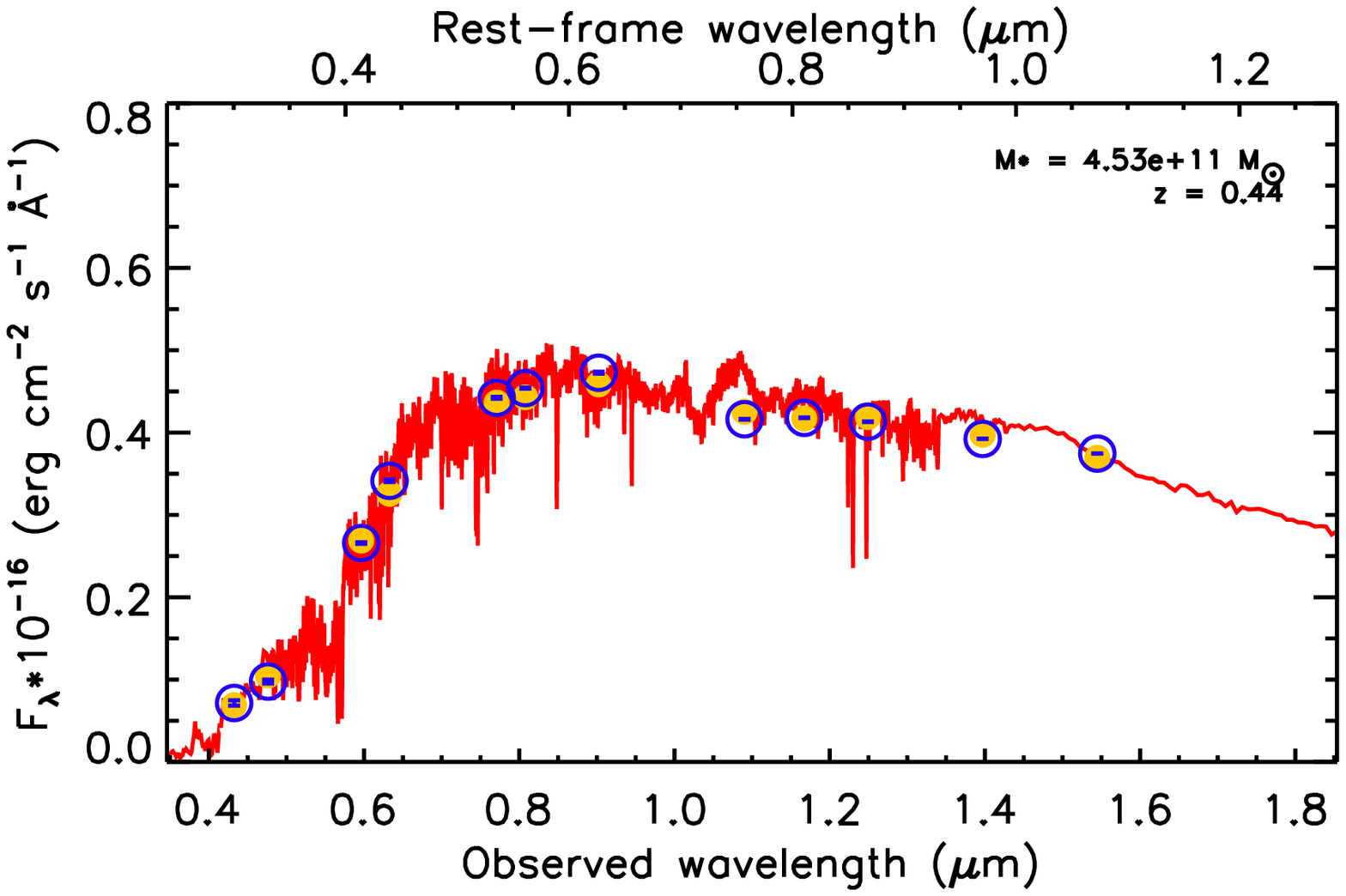}
\caption{Best-fitting composite stellar population models of the 16-band (the 4 bluest bands are not included in the plots nor in the modeling) HST photometry of the G1 (\emph{on the left}) and G2 (\emph{on the right}) lens galaxies. We use \citet{bru03} templates at solar metallicity and a Salpeter stellar IMF. Observed fluxes with 1$\sigma$ errors are represented with blue empty circles and bars, model-predicted fluxes are shown as orange filled circles.}
\label{fi06}
\end{figure*}

\begin{figure}
\centering
\includegraphics[width=0.49\textwidth]{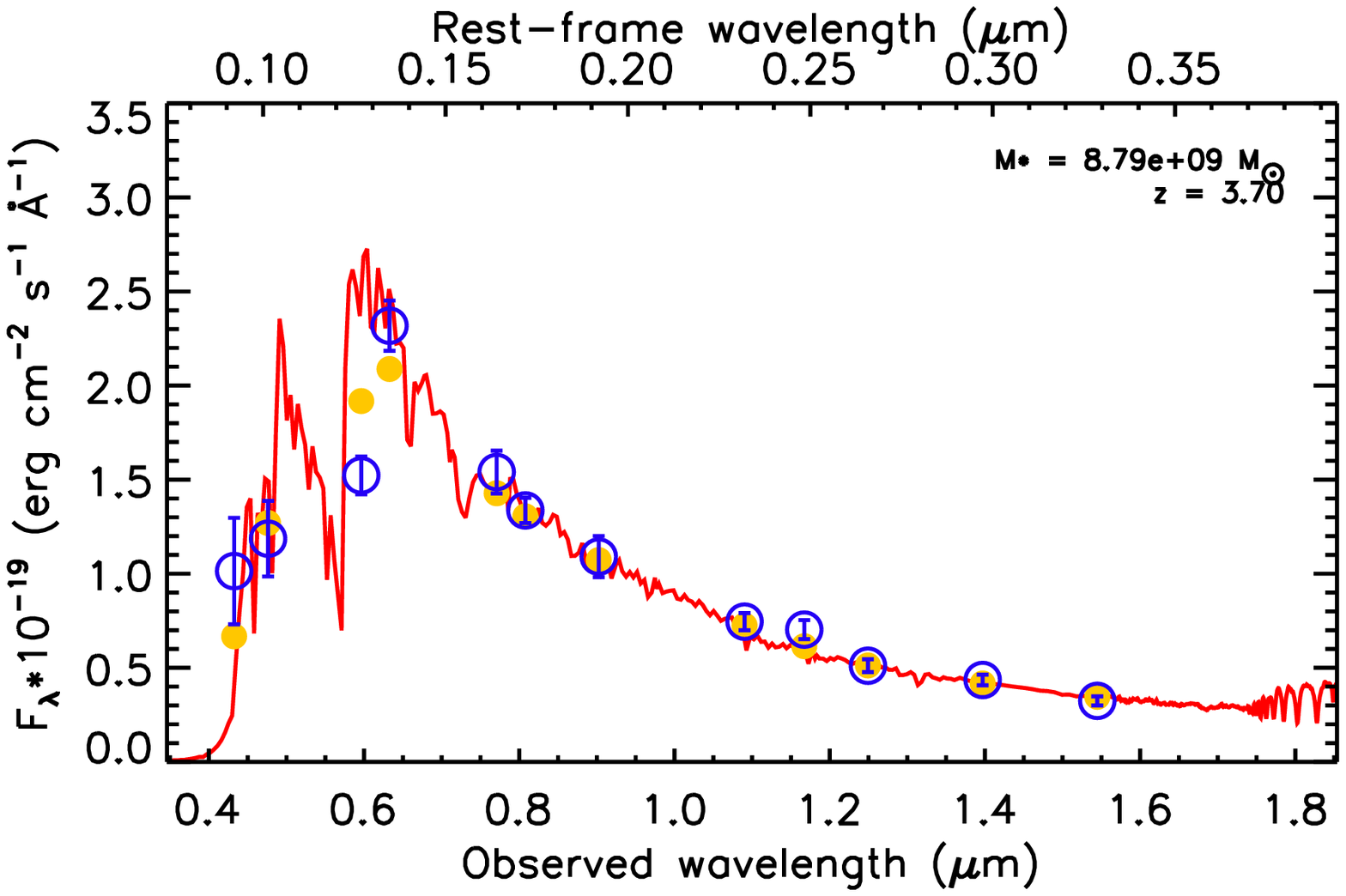}
\caption{Best-fitting composite stellar population models of the 16-band (the 4 bluest bands are not included in the plot nor in the modeling) HST photometry of the multiple images $\mathrm{A}_{5}$ and $\mathrm{B}_{5}$. We use \citet{bru03} templates at solar metallicity and a Salpeter stellar IMF. Observed fluxes with 1$\sigma$ errors are represented with blue empty circles and bars, model-predicted fluxes are shown as orange filled circles.}
\label{fi08}
\end{figure}

Here we model the multicolor photometry, composed of 16 HST bands, of the two lens galaxies G1 and G2 and of the sum of the multiple images $\mathrm{A}_{5}$ and $\mathrm{B}_{5}$. We concentrate on the measurement of the luminous mass of these objects, leaving the study of the physical properties of the source to a future work.

We use composite stellar population (CSP) models based on \citet{bru03} templates at solar metallicity and with a \citet{sal55} stellar IMF. We consider constant and delayed exponential (with a possible cut) Star Formation Histories (SFHs). We allow for the presence of dust, according to \citet{cal00}, and take into account the flux contribution of emission lines. For the two early-type galaxies and high-redshift source we choose, respectively, truncated delayed exponential and constant SFHs, which we believe are the most suitable SFHs for these classes of objects.

We summarize our modeling prescriptions and final results in Table \ref{tab7}. The best-fitting models for the two lens cluster members G1 and G2 are shown in Figure \ref{fi06} and for the lensed objects $A_{5}$ and $B_{5}$ in Figure \ref{fi08}. We have decided to exclude from the fitting and plots the 4 bluest bands because mostly affected by relevant contamination from very close objects. We have checked that removing these bands from the SED fitting does not change appreciably the results on the values of the luminous masses. In fact, photometric mass estimates are known to be more sensitive to the fluxes measured in the redder filters (e.g., \citealt{gri09}).

The best-fitting values of the luminous masses of G1, G2, and $\mathrm{A}_{5}+\mathrm{B}_{5}$ ($M_{L,G1}$, $M_{L,G2}$, and $M_{L,\mathrm{A}_{5}+\mathrm{B}_{5}}$), are, respectively, $1.7 \times 10^{10}$, $4.5 \times 10^{11}$, and $8.8 \times 10^{9}$ $M_{\odot}$. From the ranges of results obtained by considering the different photometric uncertainties, systematic errors associated to the several possible stellar population modeling assumptions (i.e., SFH, dust, emission lines), and rest-frame wavelength range covered by the HST observations, we estimate relative errors of 40\% on $M_{L,G1}$ and $M_{L,G2}$ and of 50\% on $M_{L,\mathrm{A}_{5}+\mathrm{B}_{5}}$. Furthermore, taking into account the value of the average magnification factor of approximately 9 at the positions where $\mathrm{A}_{5}$ and $\mathrm{B}_{5}$ are observed (see Figure \ref{fi01} and Table \ref{tab5}), we conclude that the measured luminous mass values and errors are $(1.7 \pm 0.7)\times10^{10}$ for G1, $(4.5 \pm 1.8)\times10^{11}$ for G2, and $(1.0 \pm 0.5)\times10^{9}$ $M_{\odot}$ for $\mathrm{A}_{5}+\mathrm{B}_{5}$.   

We remark that to avoid possible artifacts in our following investigation we have explicitly omitted the recent results (e.g., \citealt{aug10a}; \citealt{tre10}; \citealt{cap12}; \citealt{bar13}) suggesting systematic variations in the stellar IMF of a galaxy as a function of its luminous mass or stellar velocity dispersion. Taking these results into account would probably result in luminous mass estimates approximately 2 times smaller for G1 and $\mathrm{A}_{5}+\mathrm{B}_{5}$.

\section{Discussion}

\begin{table}
\centering
\caption{Values and standard deviations of the luminous mass $M_{L}$, effective velocity dispersion $\sigma_{\mathrm{SIE}}$ (or $\sigma_{*}$), and luminous over total mass fraction projected within the effective radius $f_{\mathrm{L}}(<R_{e})$ of the G1, G2, and SLACS lens galaxies, SDSS massive early-type galaxies, and some dwarf spheroidals.}
\begin{tabular}{ccccc}
\hline\hline \noalign{\smallskip}
 & $M_{L}$ & $\sigma_{\mathrm{SIE/*}}$ & $f_{\mathrm{L}}(<R_{e})$ & Ref.\\
 & ($10^{10}\,M_{\odot}$) & (km s$^{-1}$) & & \\
\noalign{\smallskip} \hline \noalign{\smallskip}
G1 & $1.7 \pm 0.7$ & $97 \pm 3$ & $0.51 \pm 0.21$ & \\
G2 & $45 \pm 18$ & $240 \pm 6$ & $0.80 \pm 0.32$ & \\
SLACS & $46 \pm 28$ & $267 \pm 39$ & & 1,3 \\
SDSS & $31 \pm 21$ & & $0.64 \pm 0.21$ & 2 \\
dSph & & $6.1 \pm 2.2$ & $0.006 \pm 0.005$ & 4,5 \\
\noalign{\smallskip} \hline
\end{tabular}
\begin{list}{}{}
\item[References --] (1) \citet{gri09}; (2) \citet{gri10}; (3) \citet{tre09}; (4) \citet{mar08}; (5) \citet{wol10}
\end{list}
\label{tab6}
\end{table}

\begin{figure}[h]
\centering
\includegraphics[width=0.44\textwidth]{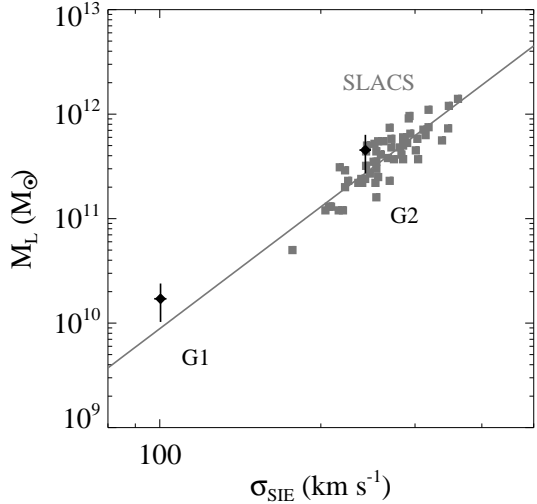}
\caption{Values of the luminous mass $M_{L}$ and effective velocity dispersion $\sigma_{\mathrm{SIE}}$ of the SLACS (grey squares) and G1 and G2 (black diamonds with 1$\sigma$ error bars) lens galaxies. The solid line shows the best-fitting line based only on the values of the lenses of the SLACS sample.}
\label{fi03}
\end{figure}

\begin{figure}[h]
\centering
\includegraphics[width=0.44\textwidth]{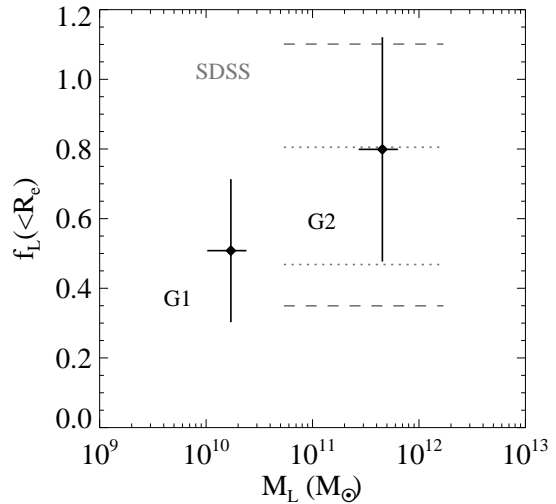}
\caption{Values with 1$\sigma$ error bars of the luminous over total mass projected within the effective radius $f_{\mathrm{L}}(<R_{e})$ and luminous mass $M_{L}$ of the G1 and G2 lens galaxies. The dotted and dashed lines show, respectively, the 68\% and 95\% confidence intervals of approximately 2$\times10^{5}$ SDSS early-type galaxies with luminous mass values between $5\times10^{10}$ and $2\times10^{12}M_{\odot}$.}
\label{fi04}
\end{figure}

In this section we compare the values of the luminous mass, effective velocity dispersion, and luminous over total mass fraction projected within the effective radius of G1 and G2 with those of three samples of SLACS lens galaxies, massive early-type galaxies from the SDSS, and dwarf spheroidals.

First, starting from the total and luminous mass estimates derived in the previous two sections, we measure for G1 and G2 the values of the fraction of luminous over total mass projected inside the effective radius, $f_{L}(<R_{e})$, in the following way:
\begin{equation}
f_{L}(<R_{e}) := \frac{M_{L}/2}{M_{T}(<R_{e})} \, .
\label{eq:05}
\end{equation}
We obtain $0.51 \pm 0.21$ and $0.80 \pm 0.32$ for G1 and G2, respectively.

Then, we consider early-type galaxies with physical properties similar to those of the lenses selected by the SLACS survey. For the SLACS galaxies, we use the luminous mass estimates by \citet{gri09}, that are obtained by fitting the galaxy SEDs with CSP models built on \citet{bru03} templates at solar metallicity and with a Salpeter stellar IMF, and the effective velocity dispersion measurements $\sigma_{\mathrm{SIE}}$, presented in \citet{tre09}, that are derived by modeling the total mass distribution of the lenses with SIE profiles. Several studies have shown that the SLACS lens galaxies are an unbiased subsample of the family of SDSS massive early-type galaxies (e.g., \citealt{bol06}; \citealt{gri10}; \citealt{aug10b}), as far as their luminous and mass properties are concerned. For this reason, we also use here the results on the luminous over total mass fractions of approximately 2$\times10^{5}$ SDSS early-type galaxies selected by \citet{gri10}. In this last study, the values of the galaxy luminous and total mass were measured under the same hypotheses adopted in this work. We show the results in Table \ref{tab6} and Figures \ref{fi03} and \ref{fi04}.

Looking at Table \ref{tab6} and Figure \ref{fi03}, we notice that G2 has values of $M_{L}$ and $\sigma_{\mathrm{SIE}}$ that are consistent with those of the galaxies in the SLACS sample. Interestingly, G1 has values of luminous mass and effective stellar velocity dispersion lower by approximately factors of 30 and 3, respectively, than the average SLACS lens galaxy, but these values are in good agreement with the extrapolation of the scaling relation based on the SLACS galaxies only. From the same Table and Figure \ref{fi04}, we observe that G2 has a value of luminous over total mass fraction that is typical of SDSS massive early-type galaxies. The value of $f_{L}(<R_{e})$ of G1 instead is smaller, but still consistent with the lower end of the distribution of the SDSS sample.

Considering the example provided by G1, we can conclude that in clusters of galaxies it is possible to study galaxy strong lensing on physical scales that are different from (i.e., smaller than) those characterizing isolated early-type galaxies (e.g., the SLACS lenses). This possibility is offered by the increase with the overdensity of the environment in the probability of one source to be strongly lensed by a (small) galaxy. In different words, the lensing cross section of a single (small) galaxy can be significantly enhanced by the presence of the mass distributions primarily on the cluster scale and secondarily on the scale of the neighboring cluster members. This opens the way to studies on the internal structure of lens galaxies over a more extended range of physical properties than done so far.

We remark that several studies (e.g., \citealt{hal07}; \citealt{lim07}; \citealt{eic13}) have found evidence for the truncation of the total mass profiles of early-type galaxies residing in galaxy clusters. Nonetheless, for a given galaxy, the value of its truncation radius is estimated to be significantly larger that that of its effective radius. This allows us to disregard, to a first approximation, the possible differences in the values of the stellar over total mass ratios projected within the effective radii of cluster and field early-type galaxies because of their different truncation radii. Moreover, we observe that the results of \citet{tre06} and \citet{gri10b} have shown that the SLACS (i.e., mainly field) and Coma (i.e., cluster) galaxies, with comparable stellar masses, do not differ appreciably as far as their inner total mass structure and stellar IMF are concerned. For these reasons, we consider appropriate to plot the values of the central luminous over total mass ratios of cluster and field galaxies in the same plots, as done in Figures \ref{fi03} and \ref{fi04}.

\begin{figure*}
\centering
\includegraphics[width=0.44\textwidth]{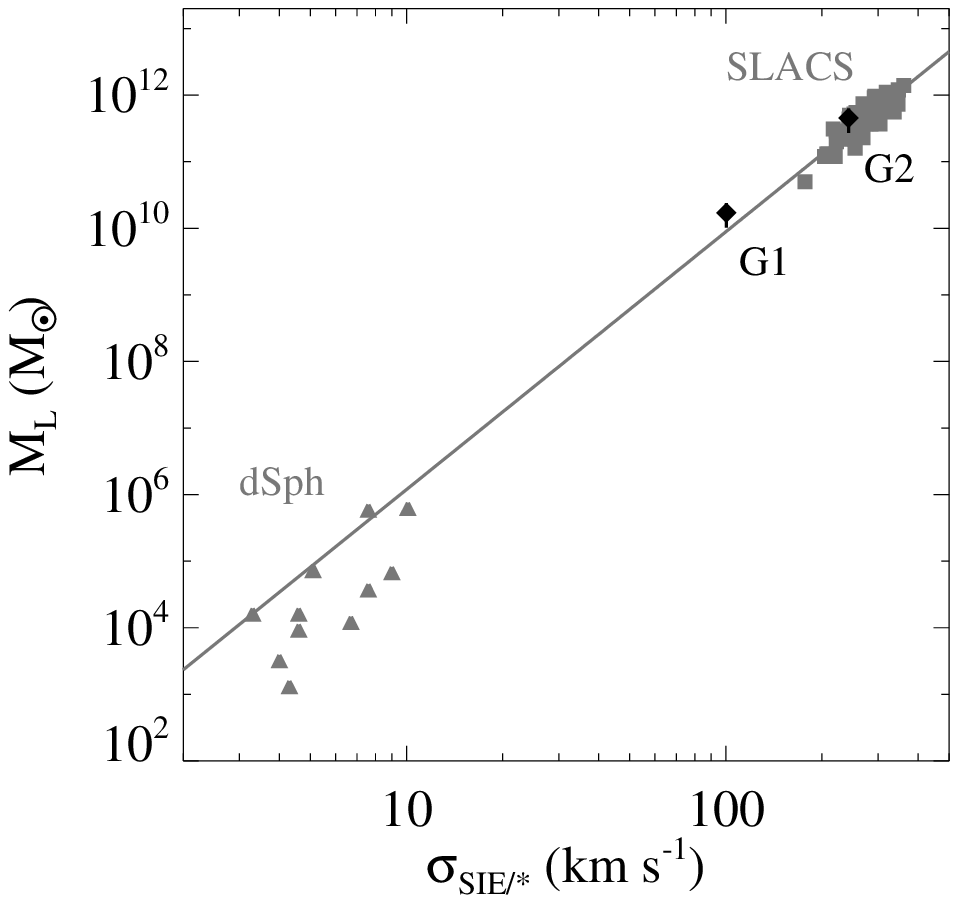}
\includegraphics[width=0.44\textwidth]{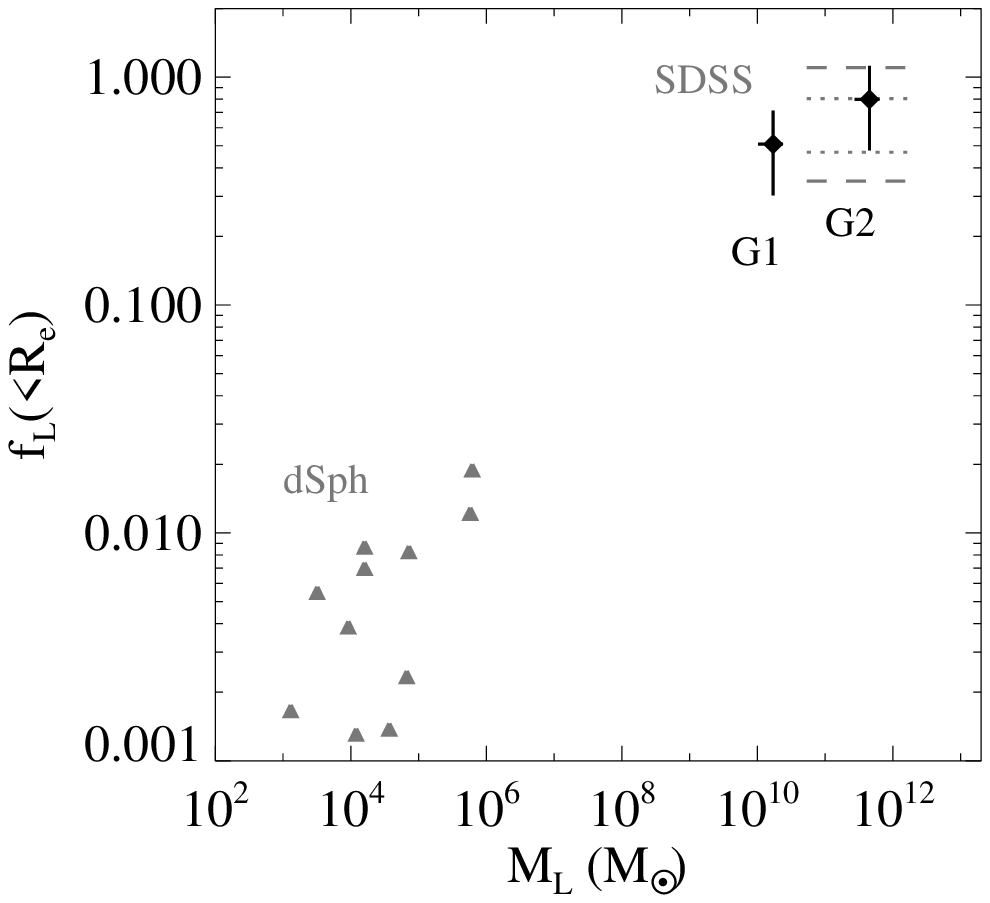}
\caption{Same quantities and symbols plotted in Figures \ref{fi03} and \ref{fi04} over ranges of values that include also 11 dwarf spheroidals (grey triangles).}
\label{fi09}
\end{figure*}

Following the previous results, we expand the intervals of physical scales plotted in Figures \ref{fi03} and \ref{fi04} to include a sample of 11 pressure-supported dwarf spheroidals (dSph) for which all the relevant quantities are available in the literature. We take the values of effective radius and luminous mass from \citet{mar08} and those of stellar velocity dispersion averaged along the line of sight $\sigma_{*}$ from \citet{wol10}. Simplistically, we decide to use the same stellar IMF (i.e., Salpeter) adopted to estimate the luminous mass of the previous galaxies and the expression given in Eq. (\ref{eq:04}) to measure the total mass projected within the effective radius. The results are plotted in Fig. \ref{fi09}.

The observed values of luminous mass and velocity dispersion of the dwarf spheroidals do not differ dramatically from the expected values obtained by extrapolating the SLACS scaling relation at several orders of magnitude difference. The projected fractions of luminous over total mass inside the effective radius show instead a clear variation from centrally luminous to dark-matter-dominated systems, moving from massive early-type galaxies to dwarf spheroidals. We speculate that the similarities and differences between these two classes of astrophysical objects might be explored effectively by extending strong lensing analyses to lenses with diverse physical scales, as started here with G1. The CLASH survey seems to be particularly well suited to this aim, as several other interesting systems of strong lensing on galaxy scale have already been discovered and are currently under investigation.

\section{Conclusions}

The combination of unprecedented HST multi-wavelength observations and VLT spectra has allowed us to perform a detailed strong lensing and stellar population analysis of an unusual system composed in total of ten multiple images of a double source, lensed by two early-type galaxies in the field of the CLASH galaxy cluster MACS 1206. Our main results can be summarized in the following points.
\begin{itemize}

\item[$\bullet$] Based on our 16-band photometry and low-resolution spectroscopy, we measure a photometric redshift of 3.7 for the source and spectroscopic redshifts of 0.436 and 0.439 for the two lens galaxies G1 and G2, respectively, thus confirming their membership to MACS 1206.

\item[$\bullet$] By modeling the total mass distribution of the cluster members and cluster in terms of singular isothermal profiles, we can reconstruct well the observed positions of the multiple images and predict a total magnification factor of approximately 50 for the source.

\item[$\bullet$] From the lensing modeling statistics, we estimate effective velocity dispersion values of $97 \pm 3$ and $240 \pm 6$ km s$^{-1}$, corresponding to total mass values projected within the effective radii of $1.7_{-0.1}^{+0.1}\times10^{10}$ and $2.8_{-0.1}^{+0.2}\times10^{11}$~$M_{\odot}$ for G1 and G2, respectively. Moreover, we obtain reasonable values for the distribution and amount of projected total mass in the galaxy cluster component.

\item[$\bullet$] Through composite stellar populations synthesis models (adopting a Salpeter stellar IMF), we infer luminous mass values of $(1.7 \pm 0.7)\times10^{10}$ and $(4.5 \pm 1.8)\times10^{11}$ $M_{\odot}$ for, respectively, G1 and G2, and $(1.0 \pm 0.5)\times10^{9}$ $M_{\odot}$ for the source, taking into account the estimated lensing magnification factor.

\item[$\bullet$] In G1 and G2, respectively, we derive luminous over total mass fractions of $0.51 \pm 0.21$ and $0.80 \pm 0.32$. We compare these values with those typical of massive early-type galaxies and dwarf spheroidals and conclude that more analyses in the CLASH fields of systems similar to that presented here will enable us to extend the investigation of the internal structure of galaxies in an important and still relatively unexplored region of the physical parameter space.

\end{itemize}

\acknowledgments
The CLASH Multi-Cycle Treasury Program is based on observations made with the NASA/ESA {\it Hubble Space Telescope}. The Space Telescope Science Institute is operated by the Association of Universities for Research in Astronomy, Inc., under NASA contract NAS 5-26555. ACS was developed under NASA Contract NAS 5-32864. This research is supported in part by NASA Grant HST-GO-12065.01-A. We thank ESO for the continuous support of the Large Programme 186.A-0798. The Dark Cosmology Centre is funded by the DNRF. We acknowledge partial support by the DFG Cluster of Excellence Origin Structure of the Universe. V.P. acknowledges the grant PRIN INAF 2010 and ``Cofinanziamento di Ateneo 2010''. The work of L.A.M. was carried out at Jet Propulsion Laboratory, California Institute of Technology, under a contract with NASA. Support for A.Z. is provided by NASA through Hubble Fellowship grant HST-HF-51334.01-A awarded by STScI. Part of this work was also supported by contract research ``Internationale Spitzenforschung II/2-6'' of the Baden W\"urttemberg Stiftung.

\clearpage




\clearpage

\clearpage

\end{document}